\documentstyle[12pt,epsfig]{article}

\title{Local law-of-the-wall in complex topography:\\ 
a confirmation from wind tunnel experiments}
\author{S.~Besio, A.~Mazzino and C.F.~Ratto\\
\small{ INFM - National Institute for the Physics of Matter,}\\ 
\small{Department of Physics, Genova University, Genova (Italy).}}
\begin{document}
\maketitle
\date{}
\vspace*{-0.6cm}
%\centerline{}
\begin{abstract}
It is well known that in a 
neutrally-stratified turbulent  flow in a deep 
constant-stress layer above a flat surface, the variation of the mean 
velocity with respect to the distance from the 
surface obeys the logarithmic law (the so-called ``law-of-the-wall'').
More recently, the same logarithmic law has been found
also in the presence of non flat surfaces. It governs the
dynamics of the mean velocity (i.e.~all the smaller scales  are
averaged out) and involves renormalized effective parameters.
Recent numerical simulations analyzed by the authors of the present Letter 
show that a more intrinsic logarithmic shape actually takes place 
also at smaller scales. Such a generalized law-of-the-wall 
involves effective parameters smoothly depending on the position
along the underlying topography.
Here, we present wind tunnel experimental evidence confirming and 
corroborating this new-found property. New results and their physical
interpretation are also presented and discussed.

\end{abstract}

Boundary layer flows, Near wall turbulence 

% 83.10.Ji  Fluid dynamics (nonlinear fluids)
% 92.60.Fm  Boundary layer structure and processes
% 47.27.Nz  Boundary layer and shear turbulence

\vspace{4mm}

\vspace{4mm}
In the realm of boundary layer flows over complex topography, 
much effort has been devoted in the last few years
to investigate both the detailed form of the surface pressure 
perturbation arising from 
the interaction between the shear flow and the underlying 
topography (see, e.g., Refs.~\cite{T81,Belch93,E90})
and its link with the effective parameters describing the large (asymptotic)
scale dynamics (see, e.g., Refs.~\cite{WM93,XT95}).
The latter regime is selected by 
observing the flow far enough from the surface and, furthermore, 
considering solely the mean velocity.
It is thus clear that with this approach all information on the 
dynamics at smaller scale  becomes completely lost.\\  
Unlike what happens for the large-scale (asymptotic) dynamics,
the description and understanding
of statistical properties of flows at `intermediate' scales
(a regime which we refer to as ``pre-asymptotic'', 
following Ref.~\cite{BMR00}) seems strongly inadequate.  
Such regime actually attracts
much attention in various applicative domains ranging from  wind
engineering
(e.g.,~for the safe design and siting of buildings), environmental
sciences (e.g., for the simulation of air pollution dispersion)
and wind energy exploitation
(e.g.,~for the selection of areas of enhanced wind speed for the economic
siting of wind turbines).\\ 
This almost unexplored regime is the main concern of the present 
Letter.
A first step in the understanding of the pre-asymptotic dynamics
has been done by the present authors in a very recent 
work \cite{BMR00}, where the analysis of 
simulations of Navier-Stokes flow fields \cite{WM93}
over two-dimensional sinusoidal topographies has been performed.
More precisely,  in the case-studies considered, 
topography takes a sinusoidal 
modulation of wavelength $\lambda$ 
(along the $x-$direction, for the sake of simplicity) and amplitude $H$,
its surface having an uniform roughness $z_0$ (with $z_0<<H$). 
Here, the dominant process governing the dynamics is 
the interaction between the shear flow and the underlying 
topography, the effect of which gives rise to a
surface pressure perturbation \cite{Belch93}. Such perturbation has
a depth of 
the order of $H$ and a downwind phase shift with respect to the
topography. The latter is the cause of
a net force on the flow, acting 
in the opposite direction of the flow itself: thus, an 
enhanced (with respect to the case of flat terrain) transfer of 
momentum towards the surface takes place.\\
Far enough from the surface,
the averaged (over the periodicity box of size $\lambda$) flow will
`see' an `effective flat surface' over which the `basic' logarithmic law
(the well known ``law-of-the-wall'' relative to flows over flat terrain
\cite{YM}) is restored but now with larger (again with respect to the
flat case) effective parameters 
${\sf u}_{\star}^{\mbox{\tiny eff}}$,
and ${\sf z}_0^{\mbox{\tiny eff}}$ \cite{WM93}, 
on account of the enhanced flux of momentum towards the surface originated 
by the aforesaid shear-flow -- topography interaction.\\
In Ref.~\cite{BMR00}, we pointed out for the first time -- as far as
we know -- that at least in the analyzed WM93 data-set \cite{WM93},
a generalized law-of-the-wall, is observed: 
\begin{equation}
U (x,z) =
\frac{u_{\star}^{\mbox{\tiny eff}}(x)}{k}
\ln \left( \frac{z}{z_0^{\mbox{\tiny eff}}(x)}\right
)\qquad\mbox{for}\qquad z>H
\;\;\; 
\label{generalizzo}
\end{equation}
where $U$ is the velocity field, $x$
is the horizontal position, $z$ is the height above the terrain,  and
$k$ is the von K\'arm\`an constant which we will take as $0.4$. Notice
that the effective parameters, $u_{\star}^{\mbox{\tiny eff}}(x)$
and $z_0^{\mbox{\tiny eff}}(x)$,
show a dependence on $x$ at scales
of the order of $\lambda$ (i.e.~the flow
`sees' some details of the topography and not only its total
cumulative effects). This is precisely the pre-asymptotic regime 
already defined in Ref.~\cite{BMR00}.\\
In the present Letter, our main goal will be to provide a first
experimental assessment confirming and corroborating the scenario
outlined in Ref.~\cite{BMR00}. In fact, the main trouble of numerical
simulations of Navier--Stokes
equations is that the impact on the results of the closure schemes,
through which small scale dynamics is accounted for, cannot be 
fully controlled \cite{WM93}. An experimental confirmation is
thus desirable.\\
To start our analysis, we briefly describe the experimental set-up
relative
to the wind tunnel experiment performed by Gong {\em et al} in 
Ref.~\cite{Gong96}.
Details on the description of the wind tunnel facility and the basic data
acquisition and analysis system are given also in
Ref.~\cite{Shokr88}.
The experiment was conducted in the AES (Atmospheric Environment Service,
Toronto, Canada) meteorological wind tunnel, which has a working volume of
$2.44\;m\;\times\;1.83\;m\;\times\;18.29\;m$ (w$\times$ h $\times$ l).
The wave model consisted
of sixteen sinusoidal waves with wavelength $\lambda\sim 610\;mm$ and
through-to-crest height $H\sim 96.5\;mm$ and was placed with its
leading edge at distance $d \sim 6.1\;m$ downstream from a honeycomb
located at
the downstream end of the contraction region. The topography can be
thus considered as a fraction of the ideal topography described by:
\begin{equation}
\label{orog}
h(x,y)=Hsin^2\left(\frac{\pi x}{\lambda}\right)
\end{equation}
where $y$ is the perpendicular-to-$x$-axis direction coordinate.\\
Two surface roughnesses were considered, corresponding to the
natural foam
surface (hereafter ``smooth case'') and to a carpet cover (``rough 
case''), respectively.
For the smooth case, velocity profile measurements gave
$z_0\sim 0.03\;mm$, 
while for the rough case $z_0\sim 0.40\;mm$. 
The flow was neutrally stratified and can be considered as a perturbation to
a stationary horizontally homogeneous infinitely
deep unidirectional constant-stress-layer flow above a plane
surface of uniform roughness, $z_0$. Thus, this basic flow
should have a logarithmic mean velocity profile,
$U(z)=(u_*/\kappa)\ln (z/z_0)$.
The values of $u_*$ were 
$\sim 0.43\;m/s$ and $\sim 0.62\;m/s$ for the smooth and the rough case,
respectively.
The free-stream velocity, $U_0$, at approximately $1\;m$ above the floor
of the tunnel, was set to about $10\;m/s$ during the measurements both in
the smooth and in the rough case. The boundary layer height, $h_B$, was
evaluated to be $\sim 600\;mm$. The rotation of the flow with the
height, produced in the numerical simulations \cite{WM93} by the Coriolis
force, is
obviously not present in the wind tunnel experiment and thus the flow is
parallel to the $x$-axis at all elevations.     

Measurements taken over the crests along the hills showed that
the flow reached an almost periodic state quite rapidly, after the $3$rd
or $4$th wave. Thus, the perturbed velocity profiles, $U(x,z)$, were
measured at selected
downstream locations between the $11$th and $12$th wave crests and a
very good agreement between the profiles over these two crests was confirmed.
This topography can be thus considered as a good approximation to a
two-dimensional topography whose shape is described by Eq.~(\ref{orog})\\
In order to compare the numerical simulations analyzed in Ref.~\cite{BMR00}
with the results from the
wind tunnel experiments here shortly described, we used the
same approach as by Finardi {\it et al.} \cite{Fin93} and Canepa {\it et
al.} \cite{Can96}. 
Accordingly, noticing that
in both cases here considered
$U_0\sim 10\;m/s$, we have kept the speeds (including the friction
velocities, $u_*$) unchanged, while the wind tunnel lengths (and times)
have been multiplied by $\lambda_W/\lambda_G\sim 1000/0.6096\sim 1640$,
where $\lambda_W$ and $\lambda_G$ are the wavelengths in the
Wood numerical simulations and in the Gong experiment, respectively. With
this change of scale, $h_B\sim 1000\;m$ and $\lambda\sim 1000\;m$ in
both cases, while the roughness lengths become $\sim 0.05\;m$ (smooth
case) and $\sim 0.66\;m$ (rough case), to be compared with the value of
$\sim 0.16\;m$ of the numerical simulations. The hill height, $H$,
becomes
$\sim 158\;m$, to be compared with the values $20\;m$, $100\;m$ and
$250\;m$ in the Wood numerical experiments.\\
\setcounter{figure}{0}
\begin{figure}[ht]
\vfill \begin{minipage}{.4\linewidth}
\begin{center}
\vspace{0.0cm}
\mbox{\psfig{figure=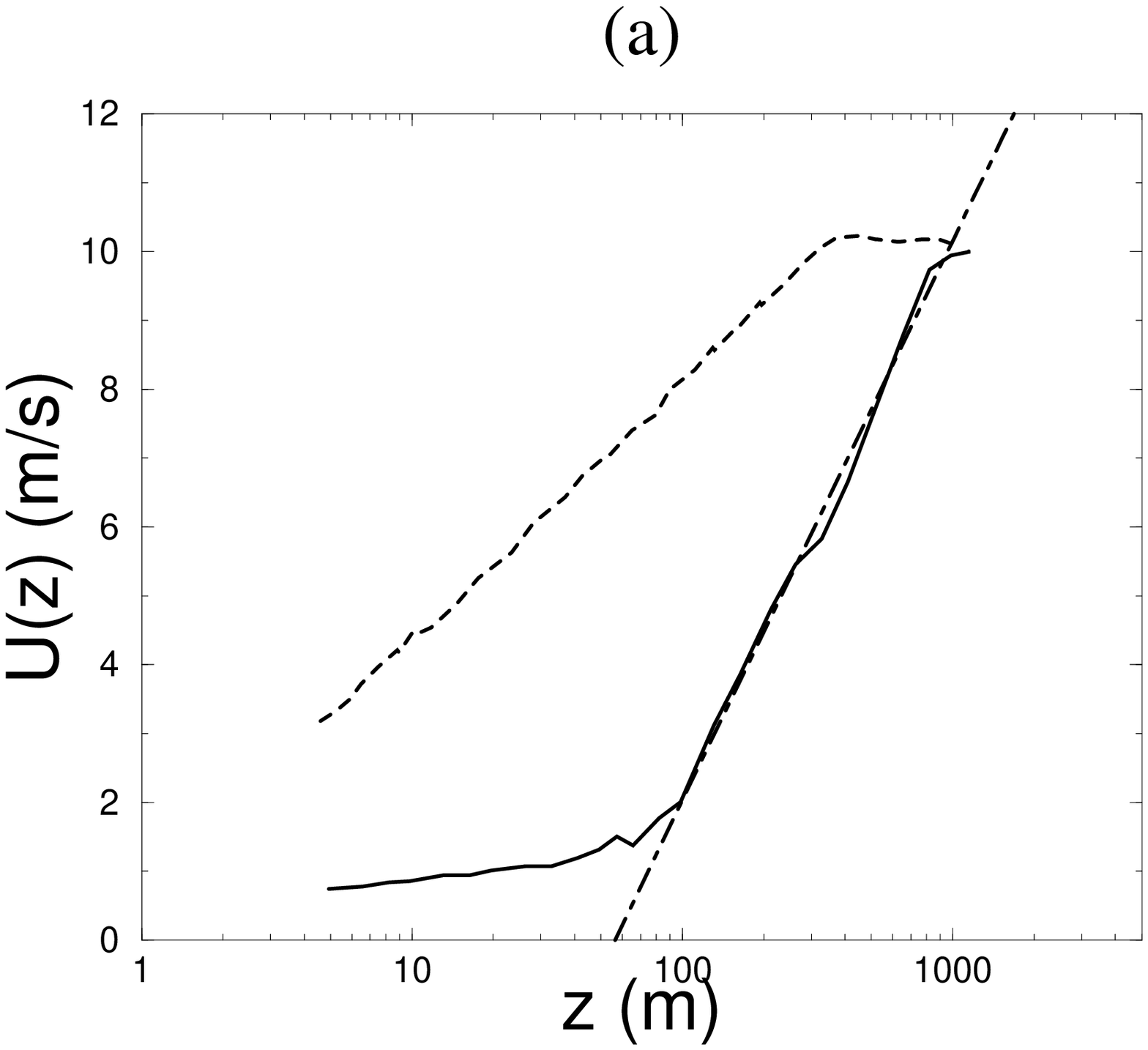,width=.9\linewidth}}
\end{center}
\end{minipage} \hfill
\begin{minipage}{.4\linewidth}
\vspace{0.0cm}
\begin{center}
\mbox{\psfig{figure=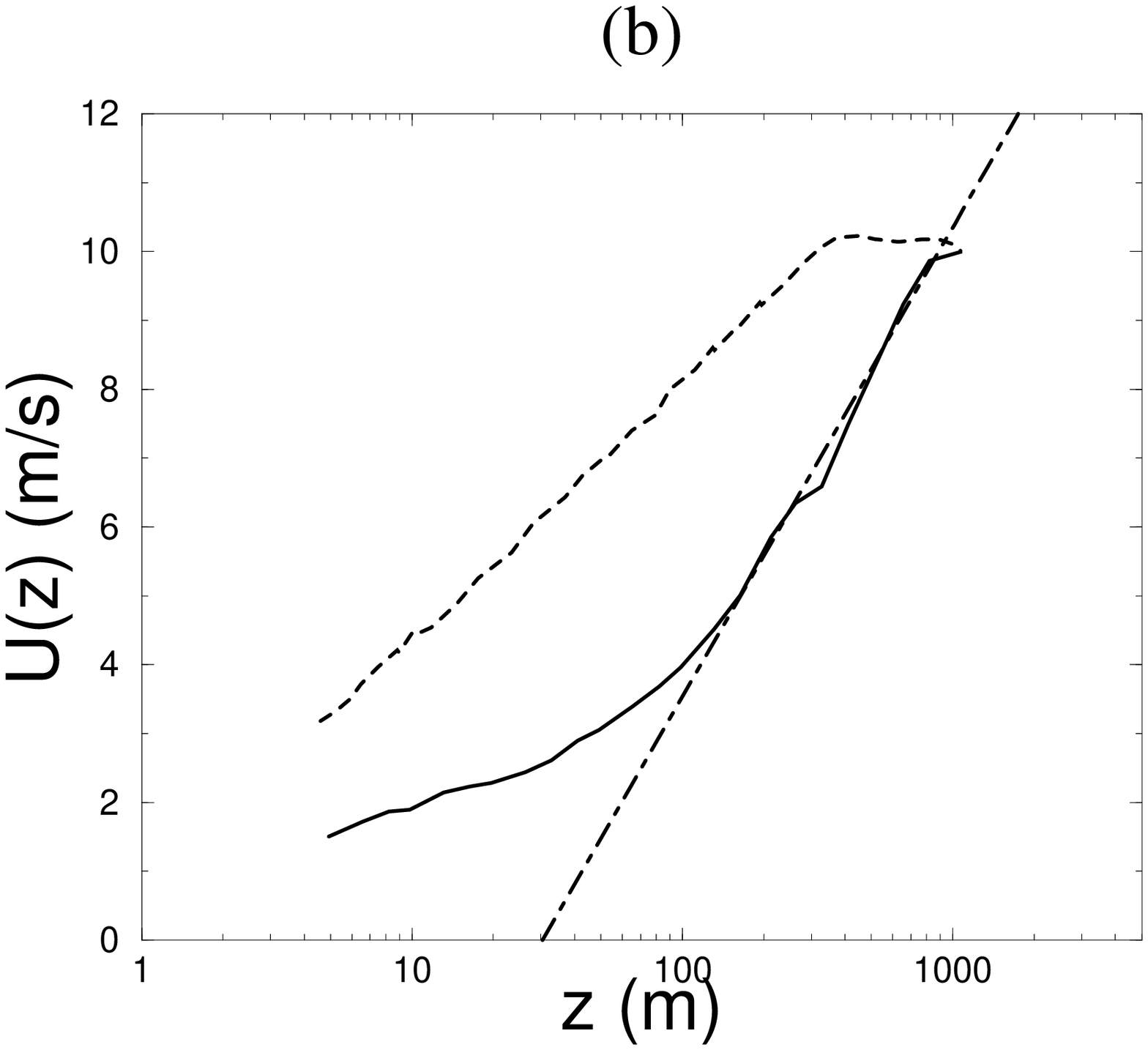,width=.9\linewidth}}
\end{center}
\end{minipage}
\vfill \begin{minipage}{.4\linewidth}
\begin{center}
\vspace{0.0cm}
\mbox{\psfig{figure=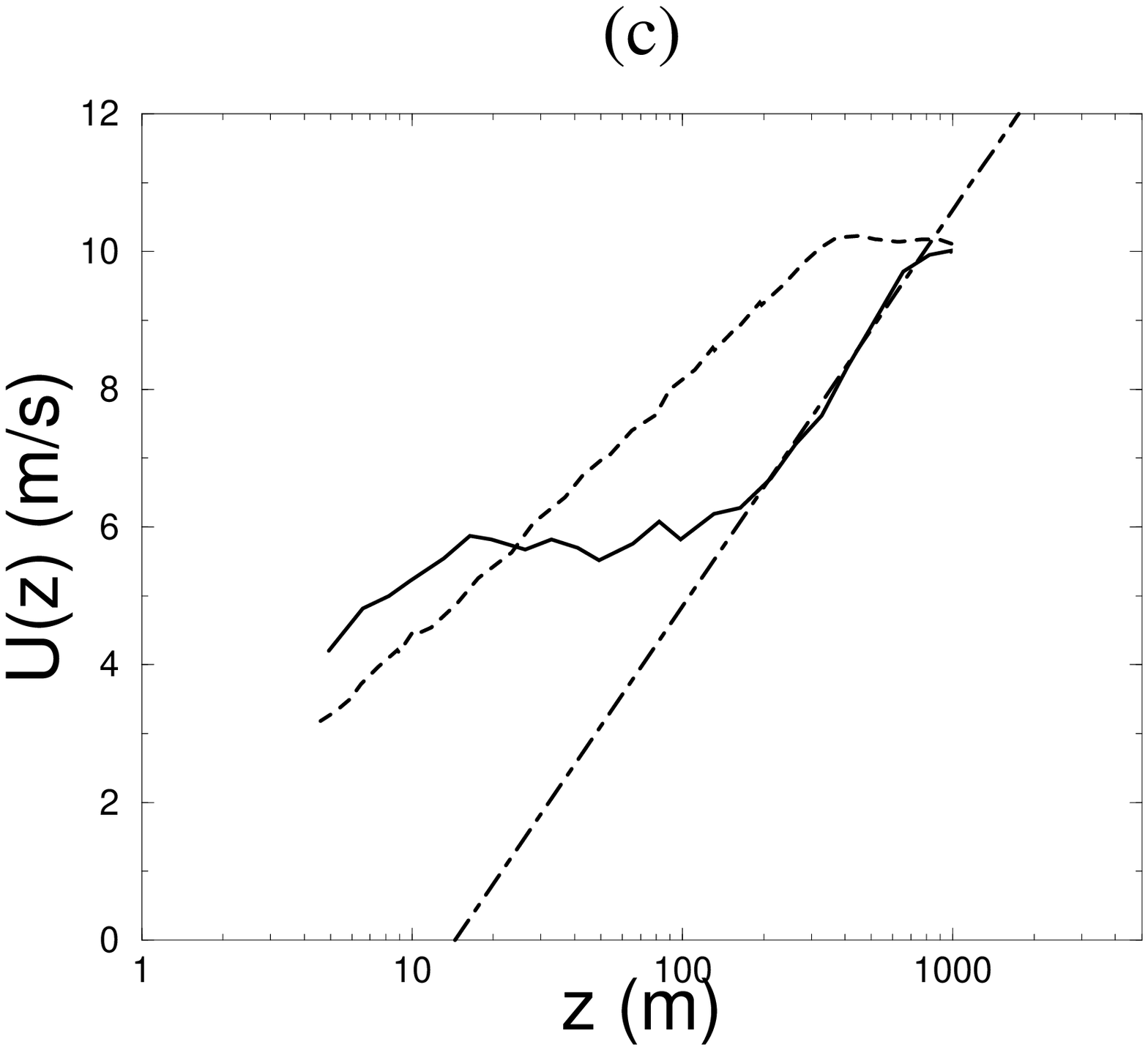,width=.9\linewidth}}
\end{center}
\end{minipage} \hfill
\begin{minipage}{.4\linewidth}
\begin{center}
\mbox{\hspace{3mm}\psfig{figure=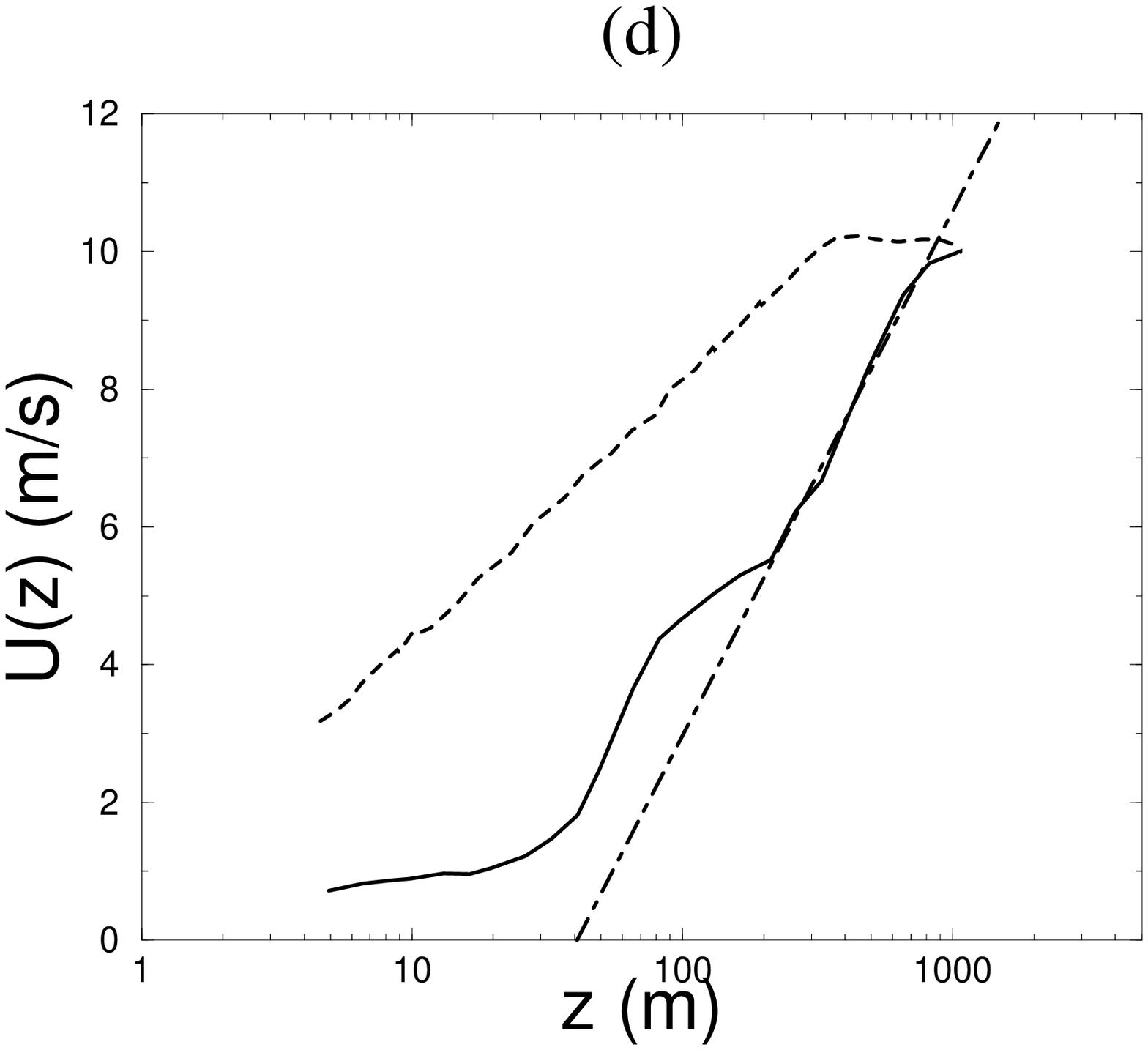,width=.9\linewidth}}
\end{center}
\end{minipage}
\vspace{-0.4cm}
\caption{The local wind speed profiles 
$U(z)$ from the wind tunnel experiment \protect\cite{Gong96}
 are plotted (solid lines) as
a function of $z$ for four different positions 
($x$ in Eqs.~\protect($1$) and \protect($2$)) 
along the hill, corresponding to (a) $x=0$, (b) 
$x=\lambda/4$, (c) $x=\lambda/2$ and (d) $x=3\lambda/4$.  
The dashed lines represent the unperturbed profile. The
dot-dashed lines represent the logarithmic law \protect(\ref{generalizzo}), 
with parameters
$u_{\star}^{\mbox{\tiny eff}}(x)$ and $z_{0}^{\mbox{\tiny eff}}(x)$
obtained by least-square fits performed inside the scaling regions.
The values of these 
 effective parameters are given 
in the text.}
\end{figure}
The first point to emphasize is that
logarithmic laws described by (\ref{generalizzo})
are evident also in the wind tunnel experiments. 
This can be easily seen in Fig.~1 
(the analogous of Fig.~1 in Ref.~\cite{BMR00}), where typical behaviours
for the horizontal wind speed profile $U(z)$ (see Eq.~(\ref{generalizzo});
for the sake of brevity, 
the dependence on the $x$-coordinate is omitted in the
notation from now on) 
as a function of $z$ are presented in lin-log coordinates for
the rough case and for four values of the $x$-coordinate corresponding to:
(a) $x=0$ (i.e.~$h=0$), (b) $x=\lambda/4$ (i.e.~$h=H/2$ upwind), (c)
$x=\lambda/2$ (i.e.~$h=H$) and (d) $x=3\lambda/4$ (i.e.~$h=H/2$ downwind),
respectively. 
Similar behaviours have been found (but not reported here for the sake
of brevity) also for the smooth case.
 From this figure, clean logarithmic region of the type described
by Eq.~(\ref{generalizzo}) are evident and both
$u_{\star}^{\mbox{\tiny eff}}(x)$ and $z_0^{\mbox{\tiny eff}}(x)$ can be
measured by least-square fits.
Specifically, for the
four above positions along the hill, we have obtained the following values
of $u_{\star}^{\mbox{\tiny eff}}(x)$ and $z_0^{\mbox{\tiny eff}}(x)$: 
$1.73\;m/s$, $0.053\;m$ (for $h=0$); $1.38\;m/s$, $0.028\;m$ (for
$h=H/2$
upwind);
$0.95\;m/s$, $0.007\;m$ (for $h=H$) and $1.27\;m/s$, $0.022\;m$ (for
$h=H/2$
downwind),
respectively. Such values can be compared with those in the absence of any
hill: $u_{\star}\sim 0.62\;m/s$ and $z_0\sim 0.66\;m$.
\begin{figure}[ht]

\vfill \begin{minipage}{.4\linewidth}
\begin{center}
\vspace{0.0cm}
%\mbox{\psfig{figure=ustar_wt.ps,width=.9\linewidth}}
\mbox{\psfig{figure=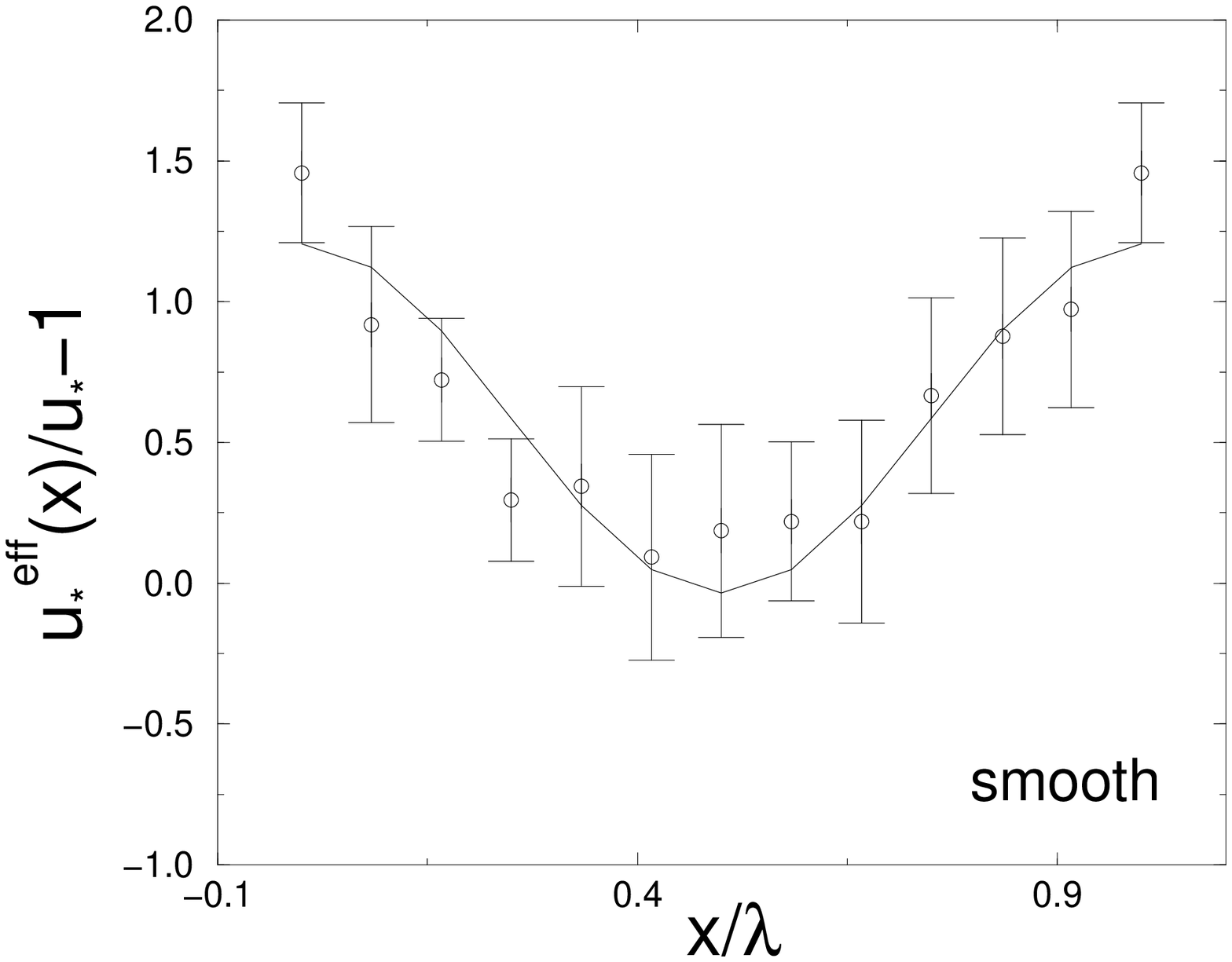,width=.9\linewidth}}
\end{center}
\end{minipage} \hfill
\begin{minipage}{.4\linewidth}
\vspace{0.0cm}
\begin{center}
%\mbox{\psfig{figure=lz0_wt2.ps,width=.9\linewidth}}
\mbox{\psfig{figure=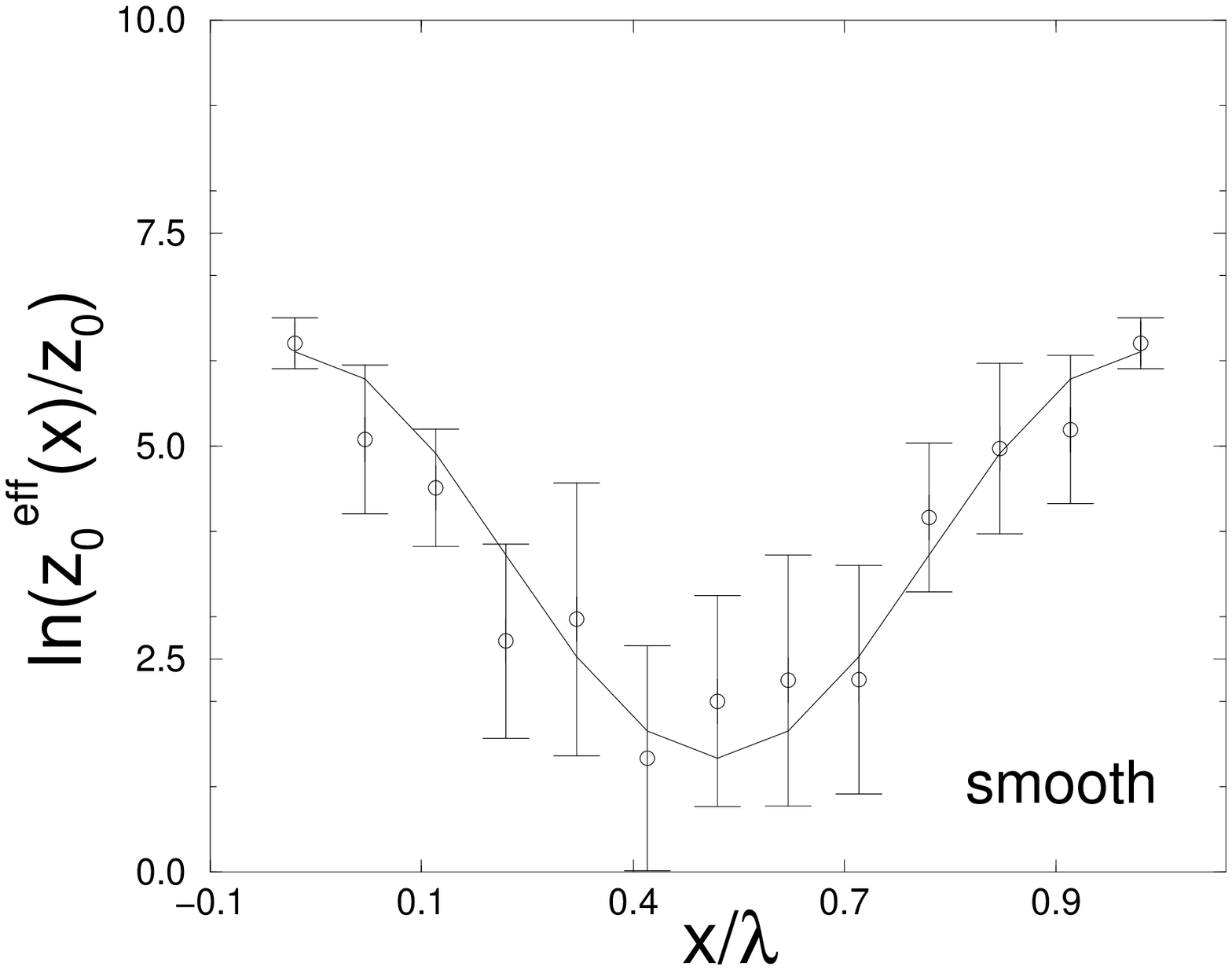,width=.9\linewidth}}
\end{center}
\end{minipage}

\vfill \begin{minipage}{.4\linewidth}
\begin{center}
\vspace{0.0cm}
%\mbox{\psfig{figure=ustar_wtr2.ps,width=.9\linewidth}}
\mbox{\psfig{figure=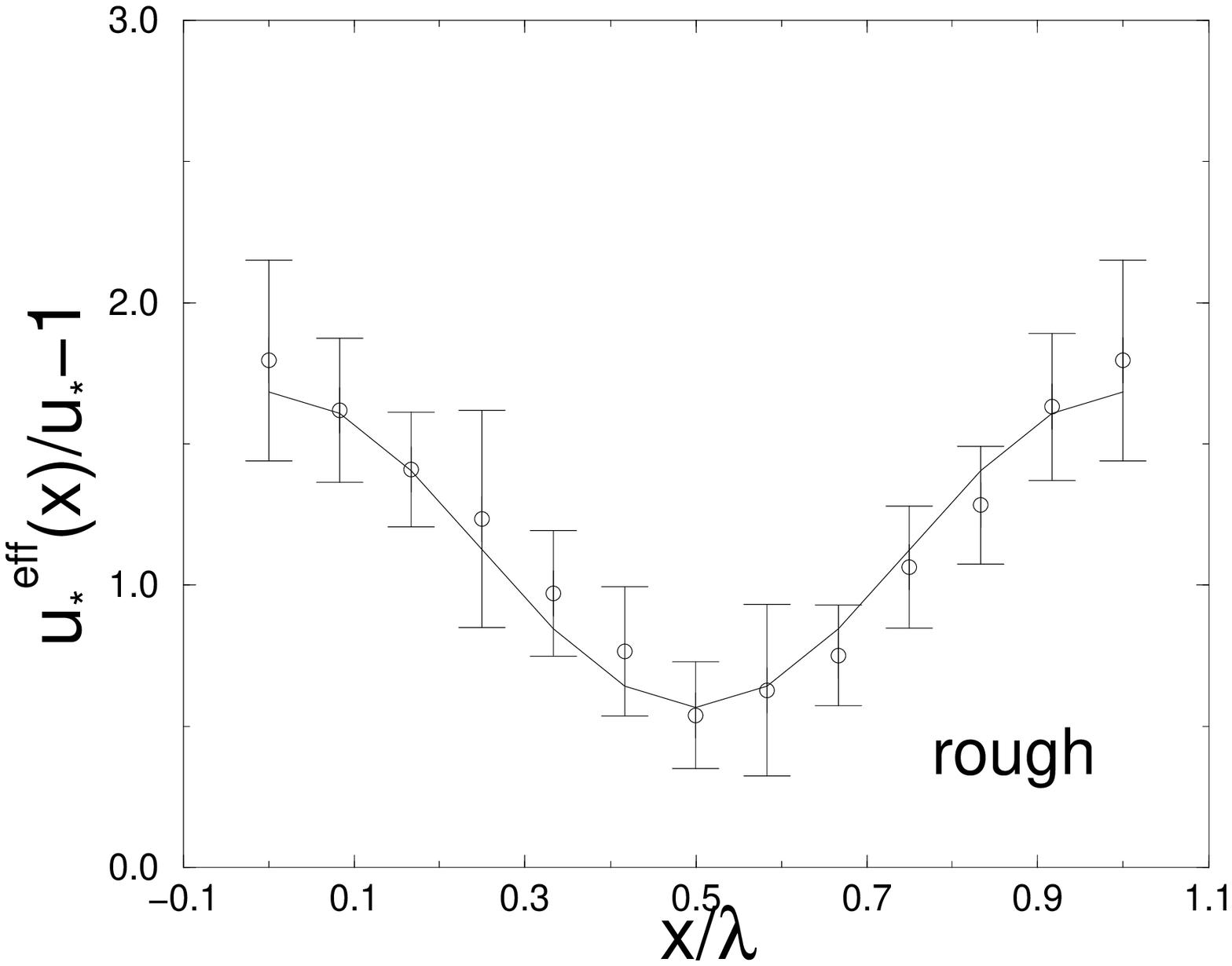,width=.9\linewidth}}
\end{center}
\end{minipage} \hfill
\begin{minipage}{.4\linewidth}
\begin{center}
%\mbox{\psfig{figure=lz0_wtr.ps,width=.9\linewidth}}
\mbox{\psfig{figure=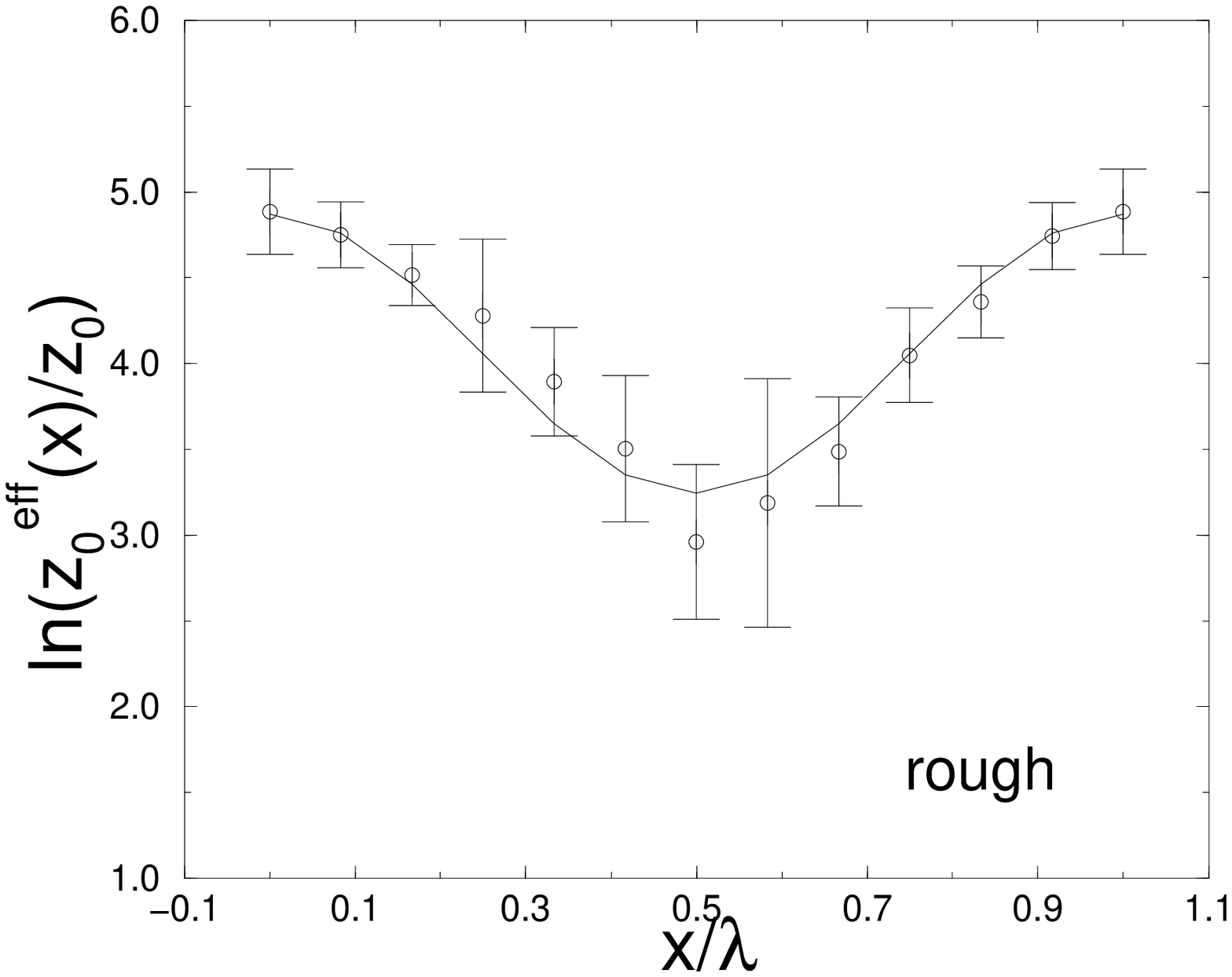,width=.9\linewidth}}
\end{center}
\end{minipage}

\caption{The measured (circles) effective parameters 
$u_{\star}^{\mbox{\tiny eff}}(x)/u_{\star}-1$ (on the left) and 
$\ln(z_0^{\mbox{\tiny eff}}(x)/z_0)$ (on the right) 
as a function of the ratio $x/\lambda$ along the
axis of the hill in the smooth case (above) and in the rough case (below).
The continuous lines represent the sinusoidal law best-fitting the
experimental data. 
%The ordinate
%on the right of each plot is relative to the hill elevation (dashed
%line).
}
\end{figure}

The results of the least-square fits are summarized in Fig.~2
where both profiles $u_{\star}^{\mbox{\tiny eff}}(x)/u_{\star}-1$ (on the
left)
and $\ln(z_0^{\mbox{\tiny eff}}(x)/z_0)$ (on the right) are shown as a
function
of $x/\lambda$ for both the smooth (above) and the rough (below) case  
(different scales in the ordinates have been adopted).
Notice that both $u_{\star}^{\mbox{\tiny eff}}(x)/u_{\star}-1$
and $\ln(z_0^{\mbox{\tiny eff}}(x)/z_0)$ have been fitted with 
the analytical expression (\ref{orog}), 
relative to the topographic profile, but with a shift of $\lambda/2$
(i.e.~$x\mapsto x+\lambda/2$ in (\ref{orog})).
More precisely, we suggest the expression: 
\begin{equation}
\label{nsformula}
y(x)=~\langle y \rangle -A_y\left[h(x)/H-1/2\right]
\end{equation}
where $y(x)$ stays for either 
$u_{\star}^{\mbox{\tiny eff}}(x)/u_{\star}-1$
or $\ln\left(z_0^{\mbox{\tiny eff}}(x)/z_0\right)$, $\langle y \rangle $ is the
average value of $y(x)$ in the interval $(0,\lambda)$ and 
$h(x)\equiv h(x,y)$ is
the topographic shape given in
Eq.~(\ref{orog}).
It should be also stressed that we have considered 
$\ln(z_0^{\mbox{\tiny eff}}(x)/z_0)$ instead of the
simpler ratio $z_0^{\mbox{\tiny eff}}(x)/z_0$, because the former
parameter is more similar 
to the topography shape than the second one.\\
It is now interesting to put together the new results here obtained
with those of Ref.~\cite{BMR00}. This allows 
to investigate at which degree of accuracy one can express 
the behaviours of quantities like the average values, $\langle y
\rangle$, and amplitudes, $A_y$, of these sinusoidal shapes solely in
terms of simple geometrical parameters. Simple considerations 
suggest to look at the ratio $H/\lambda$: this is indeed
a rough measure of the hill slope.\\ 
\begin{figure}[ht]

\vfill \begin{minipage}{.4\linewidth}
\begin{center}
\vspace{0.0cm}
%\mbox{\psfig{figure=a+bvsh_us2.ps,width=.9\linewidth}}
\mbox{\psfig{figure=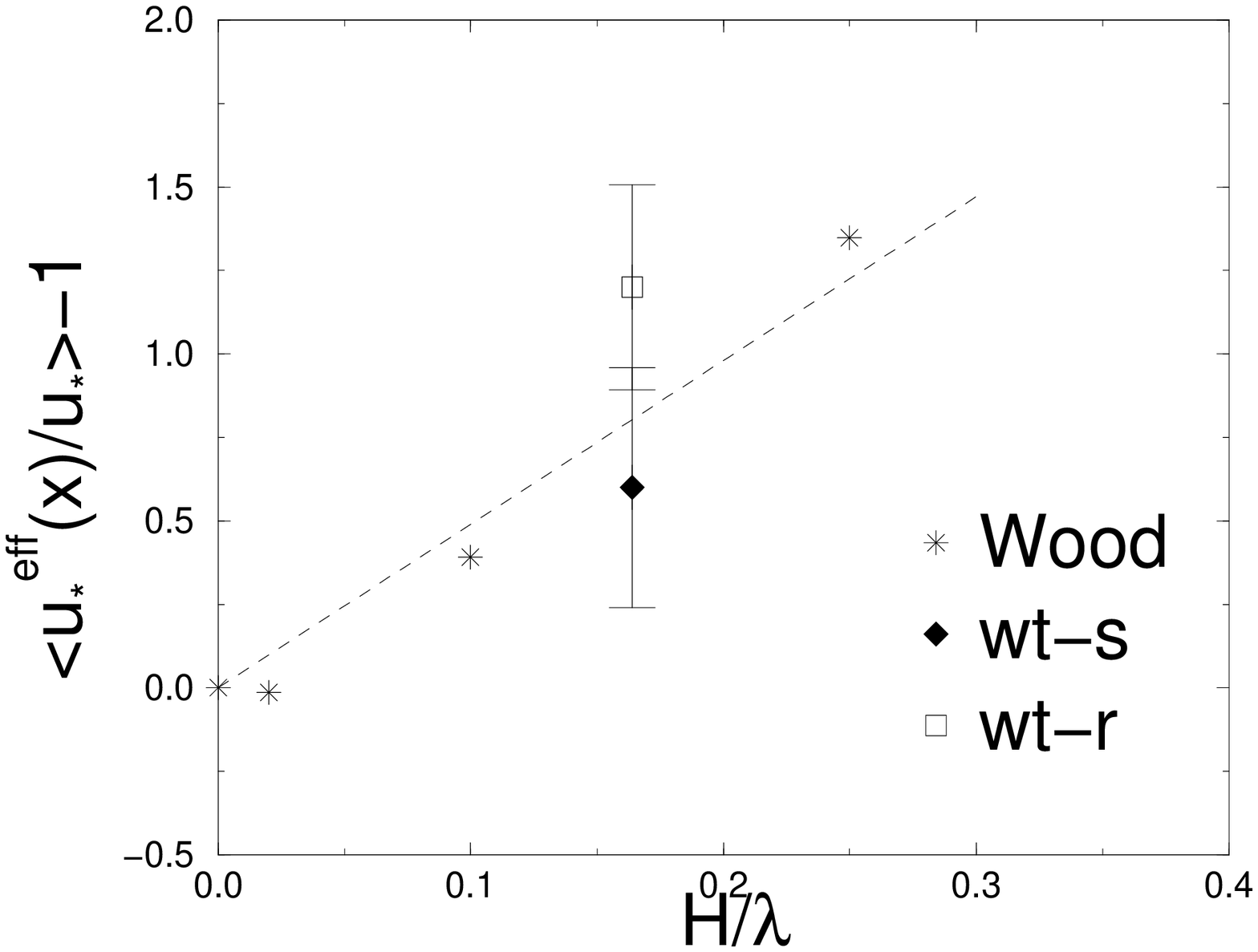,width=.9\linewidth}}
\end{center}
\end{minipage} \hfill
\begin{minipage}{.4\linewidth}
\vspace{0.0cm}
\begin{center}
%\mbox{\psfig{figure=a+bvsh_lz02.ps,width=.9\linewidth}}
\mbox{\psfig{figure=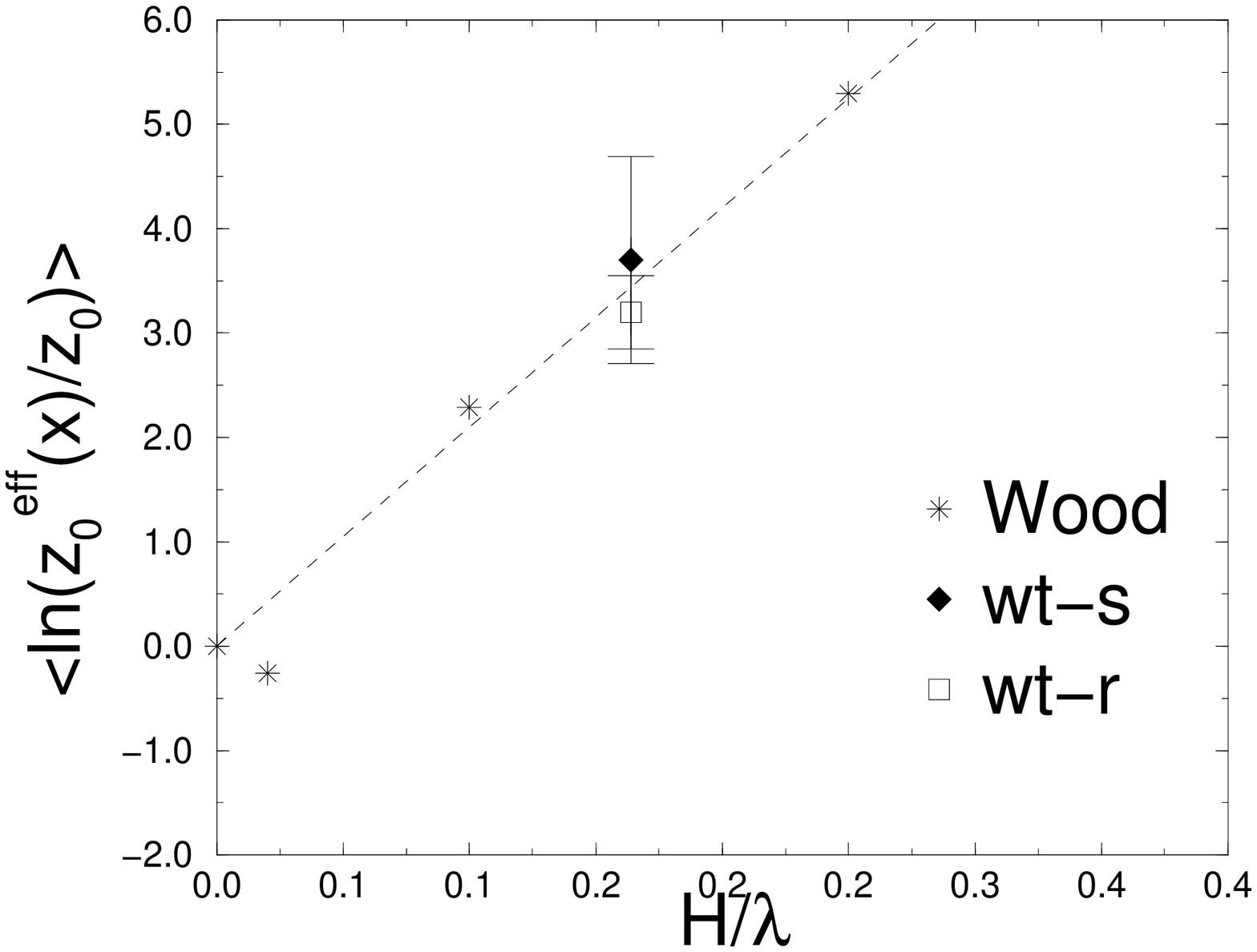,width=.9\linewidth}}
\end{center}
\end{minipage}
\caption{The mean values
$\langle u_{\star}^{\mbox{\tiny eff}}(x)/u_{\star} \rangle-1$ and
$\langle \ln(z_0^{\mbox{\tiny eff}}(x)/z_0) \rangle$ in Eq. (3) versus
$H/\lambda$. Stars
are relative to the Wood and Mason numerical simulations
($u_{\star}\sim0.44\;m/s$, $z_0\sim0.16\;m$); diamonds are
relative to the smooth case ($u_{\star}\sim0.43\;m/s$, 
$z_0\sim0.05\;m$) and squares are
relative to the rough case ($u_{\star}\sim0.62\;m/s$,
$z_0\sim0.66\;m$). Dashed lines represent the linear curve best-fitting the
Wood and Mason numerical data.} 
\end{figure} 
The values of $\langle y \rangle$ are reported in Fig.~3 for both
the smooth and the rough case. The dashed
lines (a linear fit in $H/\lambda$) are obtained considering only
the results of numerical simulations, while the values from the 
wind tunnel experiments are reported with their error bars.
These monotonic behaviours are expected
on account of the increasing of the (total) transfer of momentum
towards the surface arising for increasing slopes. 
The amplitudes $A_y$ are shown in Fig.~4. 
The dashed lines
are a parabolic fit in $H/\lambda$ and, as for $\langle y\rangle$, 
they have been 
obtained by only considering the results of the numerical
simulations. The values from the wind tunnel experiments are again
presented in the same figure.
Curves relative to the amplitudes reach a maximum for $H/\lambda\sim
0.20$, after that start to decrease. 
We can argue that two different mechanisms exist and act in 
competition.  The physical key role is played by curvature
effects \cite{T80}, already invoked in Ref.~\cite{BMR00} to explain 
the presence of minima (maxima) located above the hill top (valley) 
for both $u_{\star}^{\mbox{\tiny eff}}(x)$ and
$\ln z_0^{\mbox{\tiny eff}}(x)$.  To be more specific, let us
consider the two opposite limits $H/\lambda\ll 1$ and $H/\lambda\gg 1$,
from which we can easily isolate the two competing mechanisms.
Concerning the former limit, 
we have gentle slopes and it is well known
that in this case the flow closely follows the surface contour.
Streamlines are (weakly) curved and, as pointed out in Ref.~\cite{T80},
energy is transferred 
towards the large scale components above the hill tops,
while it blows towards the smaller scales above the valleys.
The quantity $u_{\star}^{\mbox{\tiny eff}\,2}(x)$, that is a measure 
of the energy  of turbulence, is thus smaller on the hill top
than above the valley.
Let us increase (just a little bit) $H/\lambda$. The flow
again closely follows the surface contour but streamlines 
are now more curved. As pointed out in Ref.~\cite{T80}, 
energy transfer thus increases and, as an immediate consequence,
the same happens for
the difference between the maximum and the minimum of
$u_{\star}^{\mbox{\tiny eff}}(x)$. But this means an augmentation
of its modulation amplitude.\\  
In the second limit $H/\lambda\gg 1$, 
a further important effect arises due to trapping regions
placed on the downstream hill slopes. It is in fact well known
(see, e.g.,~\cite{Milne}) that, for surface slopes large  enough,
the flow is not able to follow the contour surface and separates.
In this case, in between two hill crests, the flow is essentially trapped 
and, roughly speaking, streamlines are expunged in the wake region.
The dynamical consequence is that the flow streamlines are weakly modulated,
and this also happens for the shape of $u_{\star}^{\mbox{\tiny eff}}(x)$.
If we now decrease the ratio $H/\lambda$, trapping effects reduce 
and this means that the wake can penetrate more deeply in the
valley, with the consequent  increasing  of curvature effects
and thus of the $u_{\star}^{\mbox{\tiny eff}}(x)$ amplitude.\\
 From the inspection of these two limits, it is thus clear
that a maximum in the amplitude should be attained for a certain
finite value of $H/\lambda$, i.e.~when the two competing mechanisms
are balanced.\\
Being the maximum (minimum) of $\ln z_0^{\mbox{\tiny eff}}(x)$
directly related to the presence of the
maximum (minimum) of $u_{\star}^{\mbox{\tiny eff}}(x)$
(see Ref.~\cite{BMR00} for the discussion of this point)
the argumentations above presented hold also for 
$\ln z_0^{\mbox{\tiny eff}}(x)$.

Comparing the  values of both amplitudes
and mean values of the effective parameters extrapolated from the 
numerical simulations (i.e.~from the linear fits in Figs.~3 and 4) 
and those from the wind tunnel experiments, we notice that,
for the smooth case, experiments 
are always compatible (within the error bars) with the numerical
simulations. 
This is not always the case for the rough case. We remark that 
$u_*$ and $z_0$ (relative to the
flat terrain) are closer to the WM93 case studies in the smooth case
than in the rough case. This suggest that the expression
of both amplitudes and mean values solely in terms of geometrical quantities
like the ratio $H/\lambda$ is a reasonable approximation 
for small variations of the `bare' parameters $u_*$ and $z_0$. When
the range of variability of the latter two parameters increases,
an explicit dependence on them has to be taken into account. 

\begin{figure}[ht]

\vfill \begin{minipage}{.4\linewidth}
\begin{center}
\vspace{0.0cm}
%\mbox{\psfig{figure=bvsh_us2.ps,width=.9\linewidth}}
\mbox{\psfig{figure=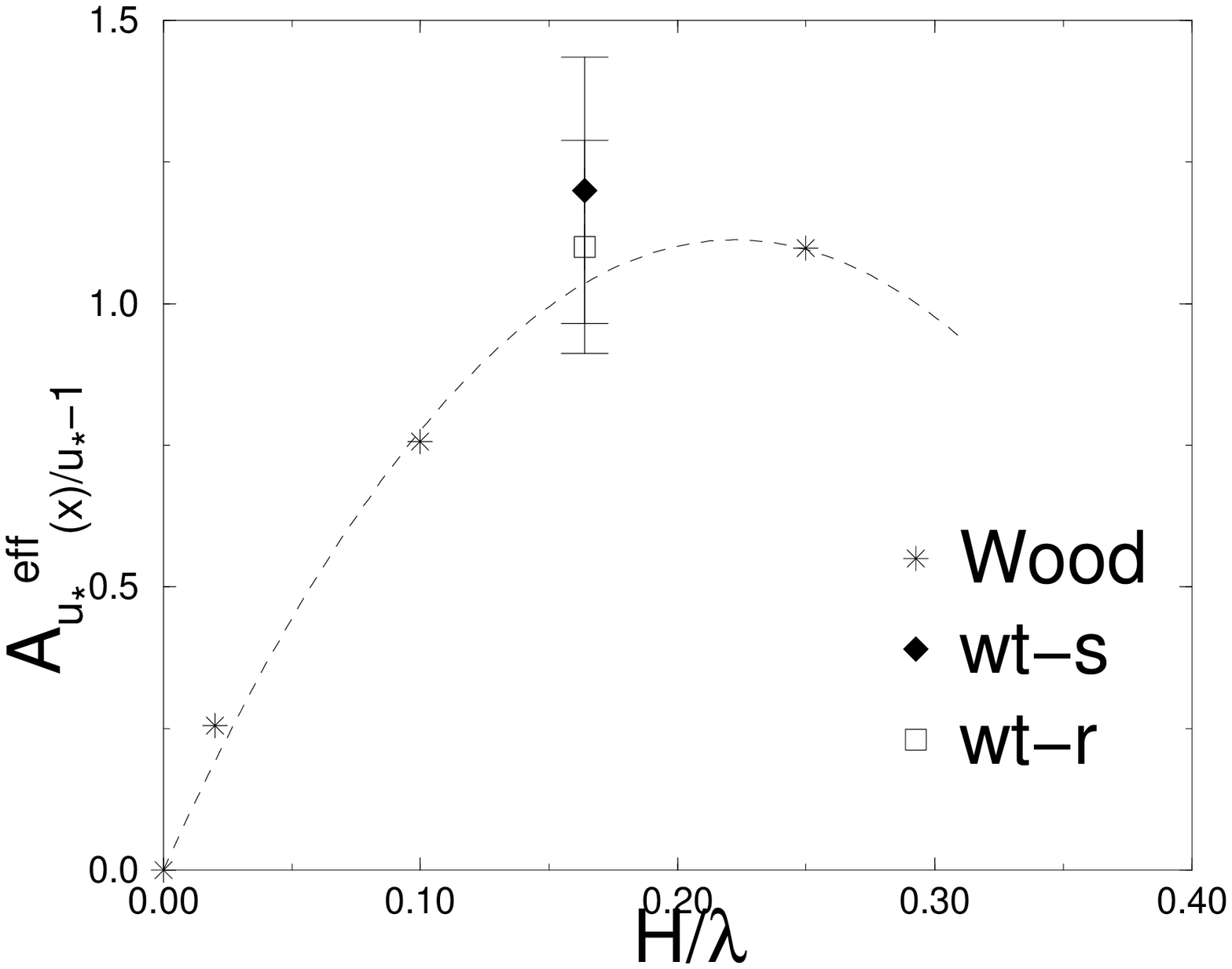,width=.9\linewidth}}
\end{center}
\end{minipage} \hfill
\begin{minipage}{.4\linewidth}
\begin{center}
%\mbox{\psfig{figure=bvsh_lz02.ps,width=.9\linewidth}}
\mbox{\psfig{figure=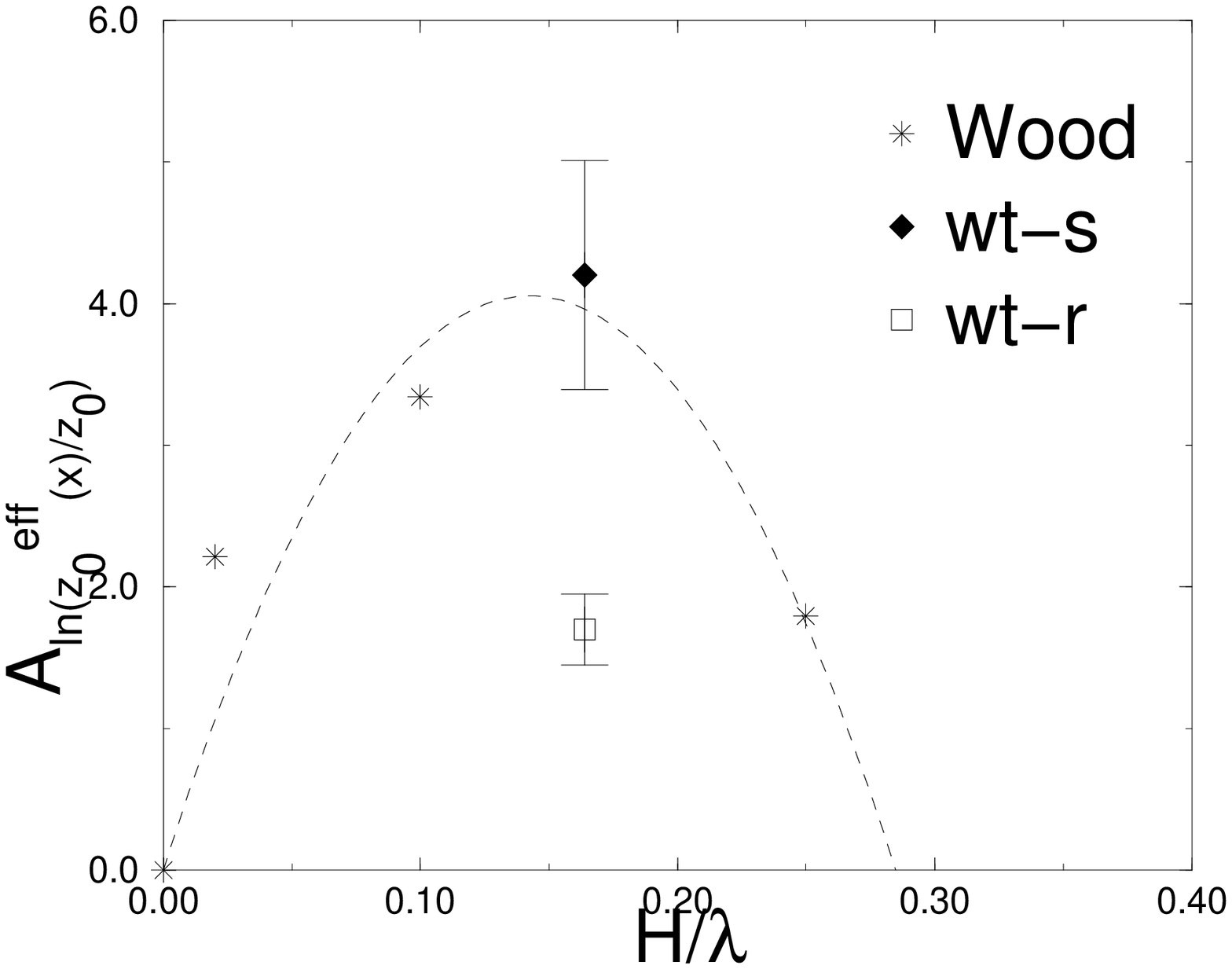,width=.9\linewidth}}
\end{center}
\end{minipage}
\caption{The values of the amplitude $A_y$ in Eq. (3) relative to
$y=u_{\star}^{\mbox{\tiny eff}}(x)/u_{\star}-1$ and $y=\ln(z_0^{\mbox{\tiny
eff}}(x)/z_0)$ versus $H/\lambda$. Stars are
relative to the Wood and Mason numerical simulations
($u_{\star}\sim0.44\;m/s$, $z_0\sim0.16\;m$), diamonds are
relative to the smooth case ($u_{\star}\sim0.43\;m/s$, 
$z_0\sim0.05\;m$) and squares are
relative to the rough case ($u_{\star}\sim0.62\;m/s$, $z_0\sim0.66\;m$).} 
\end{figure} 
 For a better evaluation and understanding of the dependence of 
$\langle y \rangle$ and $A_y$ on $H/\lambda$,  
$z_0$ and $u_*$, the analysis of more numerical and wind tunnel
experiments, and
possibly in nature, is
necessary. Nevertheless, 
the wind tunnel data here considered give a strong confirmation of 
the existence of a
pre-asymptotic regime characterized by a generalized law-of-the-wall
given by Eq.~(\ref{generalizzo}) and pointed out for the first time
in Ref.~\cite{BMR00}. Thus, this phenomenon appears as a real
physical property and not a spurious feature produced by some of the
approximations (e.g.~parameterizations of small-scale, unresolved dynamics)
used to solve the Navier-Stokes equations.\\

\vskip 0.2cm
{\bf Acknowledgements}
We are particularly grateful to P.A.~Taylor for 
providing us with his data-set relative to the 
wind tunnel experiments as well as 
many useful comments and discussions. 
Helpful discussions and suggestions by
E.~Fedorovich, D.~Mironov, G.~Solari, 
F.~Tampieri and S. Zilitinkevich are also acknowledged.

\end{document}